\documentclass[
 amsmath,
 amssymb,
 superscriptaddress
]{revtex4-2}

\usepackage{braket}
\usepackage{graphicx}
\usepackage{dcolumn}
\usepackage{bm}
\usepackage{hyperref}
\usepackage{tabularx}
\usepackage{comment}
\newcolumntype{Y}{>{\centering\arraybackslash}X}

\usepackage[mathlines]{lineno}

\usepackage{tikz}
\usetikzlibrary{arrows.meta,positioning,calc}

\renewcommand\thesubsection{\arabic{subsection}}

\usepackage{braket}
\usepackage{graphicx}
\usepackage{dcolumn}
\usepackage{bm}
\usepackage{hyperref}

\usepackage{amsmath}
\usetikzlibrary{arrows.meta,positioning,calc}

\makeatletter
\newcommand{\sublabel}[1]{%
  \label{#1}%
  \protected@write\@auxout{}{%
    \string\newlabel{#1@subonly}{{\arabic{subsection}}{\thepage}}%
  }%
}
\makeatother

\newcommand{\subnumref}[1]{\hyperref[#1]{\ref*{#1@subonly}}}

\begin{document}

\title{Adaptive nonorthogonal atomic orbitals refine charge transfer and bonding descriptors}

\author{Emily Oliphant}
 \email{eoliphan@mit.edu}
\affiliation{
 Department of Materials Science and Engineering,
University of Michigan, Ann Arbor, Michigan 48109, United States
}
\affiliation{
 Schwarzman College of Computing, Massachusetts
Institute of Technology, Cambridge, MA 02139, United States
}%
\affiliation{
 Research Laboratory of Electronics, Massachusetts
Institute of Technology, Cambridge, MA 02139
}%
\author{Emmanouil Kioupakis}
\author{Wenhao Sun}
 \email{whsun@umich.edu}
\affiliation{
 Department of Materials Science and Engineering,
University of Michigan, Ann Arbor, Michigan 48109, United States
}

\begin{abstract}
First-principles calculations implicitly encode bonding and charge transfer, but converting periodic electronic structure into interpretable and transferable descriptors remains challenging. 
Recently, we introduced Crystal Orbital Guided Iteration To atomic-Orbitals (COGITO), a tool for mapping plane-wave DFT back to an interpretable, minimal basis of adaptive atomic orbitals. 
Here, we show how the high-dimensional bonding information from COGITO can be reduced into chemically useful descriptors of covalent bonding, charge transfer, and crystal-dependent atomic chemistry without collapsing the real-space origin of the interactions. 
Compared to methods that use static or orthogonal orbital bases, COGITO descriptors of short- and long-range bonding show more stability and a better ability to distinguish between ionic, covalent, and metallic bonding regimes. 
Lastly, COGITO reliably decomposes band dispersion into range-resolved contributions, allowing us  to verify and add to the orbital origins of silicon’s low-symmetry conduction band minimum.
Overall, the adaptive COGITO-based descriptors encode electronic structure, offering a well-structured route to expose the orbital interactions and charge transfer responsible for electronic properties.

\end{abstract}

\maketitle


\section*{Introduction}

Interpretable descriptions of chemistry and structure provide the scaffolding for rationalizations of materials properties. 
At their best, simple concepts such as electronegativity difference or bond geometry offer satisfying explanations for property trends such as crystal structure selection\cite{Goldschmidt1926,Mooser_1959,Chelikowsky_1986} or hardness\cite{Gao_hardness_2003,Li_hardness_2012,Guo_hardness_2008}. 
Quantitative prediction, however, often requires additional complexity. 
As such, materials informatics has relied on combinations of atom features and local environment features to better anticipate a compound's properties.\cite{Bartok_soap_2013,Bartel2019,Choudhary_machine_2018}
While prediction accuracy can be improved with a variety of machine learning approaches\cite{Ward2016,Dunn2020}, a model built on predetermined descriptors must infer 
electronic structure changes from geometry and composition rather than being supplied with it directly.

To better represent the action of quantum mechanics, the atom features and local environment features should be allowed to adapt in response to each other. 
Providing an algorithmic mechanism for this missing adaptability is a key reason why machine learning architectures based on graph neural networks are so successful. \cite{Xie_CGCNN_2018,Gilmer_MPNN_2017,Chen_MEGNet_2019}
However, learned representations are necessarily \textit{post hoc} explanations of a particular trained model rather than explicit representations of electronic structure.
In this article, we present a route to explicitly recover how local atoms and bonds mutually adapt from Kohn-Sham Density Functional Theory (KS-DFT). 
Here, we establish the reliability of these descriptors through their numerical robustness and fidelity to known chemistry.
In contrast, we show where non-adaptive or orthogonal methods to obtain quantum-based descriptors fall short of achieving reliability requirements.

Previous ventures to obtain bonding descriptors directly from DFT calculations are limited by how consistently and without loss the periodic Kohn–Sham wavefunctions are mapped back to atomic features and an inability to separate range (\textbf{R}-space) in periodic (\textbf{k}-space) quantities. Commonly, the KS wavefunctions are projected onto a predetermined atomic orbital basis to inform band character, charge transfer, and orbital bonding. While limited in quantitative and chemical accuracy by the quality of the initial basis, the projected matrices can be used to transform any operator (e.g. density, overlap, and Hamiltonian) from the KS orbital basis to the projected atomic orbital basis (detailed in Methods~\ref{sec:local_ham}). Such metrics have been used as descriptors with mixed success.\cite{Naik_critical_2026} 
We suggest that part of this inconsistency originates in the orbital basis itself, and therefore focus here on establishing descriptor reliability before deploying these descriptors for property prediction.

\newpage
A crucial limitation for extracting transferable descriptors is that the Kohn–Sham states in periodic systems are resolved by a momentum quantum number \textbf{k}. Purely local quantities cannot be obtained from individual \textbf{k}-space quantities since it requires projecting \textit{Bloch} atomic orbitals, which are periodic images of the local atomic orbitals. While some relevant one-center terms (e.g. Mulliken population analysis) are invariant under a switch from local to Bloch states, the periodic images convolve directional origins, prevent clear range-resolving quantities, and limit direct chemical interpretation of other one-center terms and all two-center terms.

In Wannier-based methods, the interactions between local orbitals are recovered in \textbf{R}-resolved orbital overlaps and Hamiltonians by doing a Fourier transform over reciprocal \textbf{k}-space quantities.\cite{marzari_maximally_2012} 
Additionally, Wannier orbitals adapt to the KS states, providing a path to naturally capture how the predefined orbitals adjust to different local environments.
Yet, while Wannier orbital Hamiltonians are widely used for interpolation and model construction, their full information content is rarely exploited for chemical analysis. \footnote{Implementations of Wannier-based CO*P (WO*P)\cite{WOOP_WOPP,Terakura_cov_polar, kundu_population_2021} have either focused on polarization (WOPP) or did not fully develop the mathematical and computational structure that enables easy resolution over bond direction and range.}
Part of this is because atomic-like Wannier orbitals face a substantial problem for interpretability: the Wannier functions (and matrices which are in their basis) deviate substantially from atomic orbitals by mixing with other orbitals on the central and neighboring atoms.\cite{qian_quasiatomic_2008} These deviations arise from a combination of orthogonality, orbital mixing, and fixed-overlap constraints.\cite{Oliphant_COGITO_2026}

Our new method Crystal Orbital Guided Iteration To atomic-Orbitals (COGITO)\cite{Oliphant_COGITO_2026} resolves previous limitations from static projections and orthogonal Wannierizations by obtaining a highly atomic nonorthogonal Wannier representation of Kohn–Sham electronic structure. 
COGITO constructs strictly atomic orbitals that span the selected Kohn–Sham states while allowing the radial form and overlap of the basis to adapt to the chemical environment.
The resulting representation preserves the atomic character while retaining sufficient quantitative and chemical accuracy for tight-binding interpolation and bonding analysis. 
Across a set of 200 materials, the COGITO basis achieves a median charge spilling of 0.3\%, median orbital mixing of 0.6\%, and tight-binding interpolation with a median valence band distance of 1.3 meV. 
In \textbf{Results~\ref{sec:cog_update}}, we outline how the COGITO algorithm is modified to improve the numerical stability of COGITO descriptors and how to extract the full dimensionality of bonding data.



In three examples, we demonstrate how our descriptor framework unlocks range-resolved bonding and reliable crystal-dependent atomic chemistry. 
In tandem, we clarify the importance for a simultaneously adaptable and atomic orbital basis by comparing COGITO results with PAW pseudo-orbitals, MLWFs, or LOBSTER. 
In \textbf{Results~\ref{sec:bonds_variety}} and \textbf{~\ref{sec:bonds_gan}} we show how COGITO captures differences between ionic, covalent, and metallic bonding regimes and stabilizes short- and long-range bonding in a variety of compounds. In \textbf{Results~\ref{sec:bonds_si}}, we demonstrate how COGITO provides clear insight on which range-resolved bonds shape electronic band dispersion. 
Together, these examples show how adaptive atomic-orbital representations provide a general route for converting first-principles calculations into chemically interpretable quantum-bond descriptors. 


\section*{Results}

\subsection{Building transferable, high-dimensional descriptors}
\label{sec:cog_update}

Here, we provide a brief summary of the mathematical process COGITO uses to construct local matrices (detailed in Methods~\ref{sec:local_ham}), summarize key improvements for numerical consistency that ensure transferability (detailed in Methods~\ref{sec:cog_robust}), and outline the mathematical structure of bonding analysis.

COGITO maps plane-wave Kohn--Sham (KS) states onto a minimal basis of strictly atomic, nonorthogonal orbitals whose radial forms and overlaps adapt to the chemical environment.\cite{Oliphant_COGITO_2026} 
Projecting the KS states onto Bloch sums of these orbitals yields \textbf{k}-dependent Hamiltonian and overlap matrices.
A discrete Fourier transform then converts these into the real-space matrices \(H_{\alpha\beta}^{\mathbf{R}}\) and \(S_{\alpha\beta}^{\mathbf{R}}\), which couple orbital \(\alpha\) in the home cell to orbital \(\beta\) in the cell translated by \textbf{R}.
Every descriptor in this work is built from these local matrices.
If only individual \textbf{k}-dependent matrices are used, interactions with different translated images of the same atom fold into a single matrix element, making them inseparable. See Methods~\ref{sec:local_ham} for more detail.


To serve as descriptors, these matrices require a basis that changes between compounds only when the chemistry does, and not with cell size, grid resolution, or numerical noise.
We therefore made two changes to COGITO, detailed in Methods~\ref{sec:cog_robust}.
First, the atomic orbitals are now fit directly to the Wannier orbitals. Previously they were approximated from the \textbf{k}=\textbf{0} Bloch orbitals, an approximation whose quality depended on the size of the unit cell.
Second, we introduce a radial function restricted to smoothly decaying orbitals that remains flexible enough to contract or expand with charge transfer and local environment (Eqn.~\ref{eq:cohp9d}).
When fit to noised Pb 6\textit{s}, 6\textit{p}, and 5\textit{d} pseudo-orbitals over 25 noise seeds, the new form reduces the mean relative standard deviation of the fitted parameters to 0.105, compared with 0.561 for a sum of three Gaussians and 0.831 for 16 spherical Bessel functions.
As a final test with both improvements, we check convergence of downstream COGITO properties with Wannier grid spacing on four GaN polymorphs and with supercell size in sphalerite. We find after converging to 0.25~Å, the residual grid dependence is over an order of magnitude smaller than the differences between polymorphs.

With the local Hamiltonian, Crystal Orbital Hamilton Population (COHP)\cite{dronskowski_crystal_1993} can be resolved over every index:
\begin{equation}
\mathrm{COHP}_{\alpha\beta\mathbf{R}}^{n\mathbf{k}}
= {c_{\alpha n}^{\mathbf{k}}}^{*}c_{\beta n}^{\mathbf{k}}\,H_{\alpha\beta}^{\mathbf{R}}\,
e^{i\mathbf{k}\cdot\left(\mathbf{R}-\boldsymbol{\tau}_{\alpha}+\boldsymbol{\tau}_{\beta}\right)},
\label{eq:cohp9d}
\end{equation}
where \(c_{\alpha n}^{\mathbf{k}}\) is the coefficient of orbital \(\alpha\) in band \(n\), obtained from the generalized eigenvalue problem of the \textbf{k}-dependent Hamiltonian and overlap.
The result is nine-dimensional, spanning band \(n\), the three components of \textbf{k}, the orbital pair \(\alpha\beta\), and the three components of \textbf{R}. It retains the energy, momentum, orbital, and real-space origin of every interaction.
Replacing \(H_{\alpha\beta}^{\mathbf{R}}\) with \(S_{\alpha\beta}^{\mathbf{R}}\) gives the analogous Crystal Orbital Overlap Population (COOP), which counts electrons rather than energies.
When COHP is instead constructed from projected \textbf{k}-dependent Hamiltonians alone, the sum over \textbf{R} is implicit, and in small primitive cells distinct neighbor shells collapse onto the same matrix element.
In diamond-structure silicon, for example, second nearest-neighbors fold into the onsite terms and third nearest-neighbors into the first (\textbf{Results~\ref{sec:bonds_si}}).

Additionally, because \(H_{\alpha\beta}^{\mathbf{R}}\) and \(S_{\alpha\beta}^{\mathbf{R}}\) form a compact tight-binding model, COHP on any \textbf{k}-path or dense \textbf{k}-grid follows by Fourier interpolation without recalculating KS wavefunctions. And since COGITO leverages all crystal symmetry, only a standard self-consistent calculation on an irreducible \textbf{k}-grid is required.

Each descriptor used in this work is a reduction of \textbf{Eqn.~\ref{eq:cohp9d}} (Methods~\ref{sec:cohp_form}).
Summing over occupied bands and \textbf{k}-points gives the integrated COHP (iCOHP) of each orbital pair and translation.
Summing further over the orbitals of atoms A and B gives the atom-pair bond energy \(\mathrm{iCOHP}_{AB\mathbf{R}}\) and bond charge \(\mathrm{iCOOP}_{AB\mathbf{R}}\), which set the bond cylinders in the crystal bond diagrams of \textbf{Figs.~\ref{fig:vis_five}} and \textbf{\ref{fig:gan}}.
Because every interaction retains its interatomic distance, bonds can be filtered by range. We filter them into short- and long-range bond charges (\textbf{Results~\ref{sec:bonds_variety}}) and by neighbor shell in the silicon band structure (\textbf{Results~\ref{sec:bonds_si}}).
The onsite (A=B, \textbf{R}=\textbf{0}) iCOOP gives the electrons residing on each atom, while summing each orbital's iCOOP over all partners gives charge-complete Mulliken populations. The resulting Mulliken charges supply the Madelung energies in \textbf{Results~\ref{sec:bonds_gan}}.

\subsection{Range-resolved bond descriptors across bonding regimes}
\label{sec:bonds_variety}
First, we analyze bonding in the set of nonmagnetic materials compiled by Vitale et al. \cite{vitale_automated_2020}, which spans the spectrum of covalency, ionicity, and metallicity. An effective atomic basis should not only reproduce the KS wavefunctions with realistic measures of covalency and ionicity, but also resolve fundamentally different bonding regimes into descriptors that track the underlying physics of chemical bonding. We test this in three ways. First, we examine representative binaries that illustrate the covalent, ionic, metallic, and mixed bonding regimes. Second, we group binaries by the element classes of their two constituents and examine how each group distributes across short- and long-range bonding. Third, we perform supervised classification of the compound groups and analyze how separability between groups changes when switching from VASP PAW pseudo-orbitals to COGITO. 

\begin{figure*}[hbt!]
\centering
\includegraphics[width=6.3in]{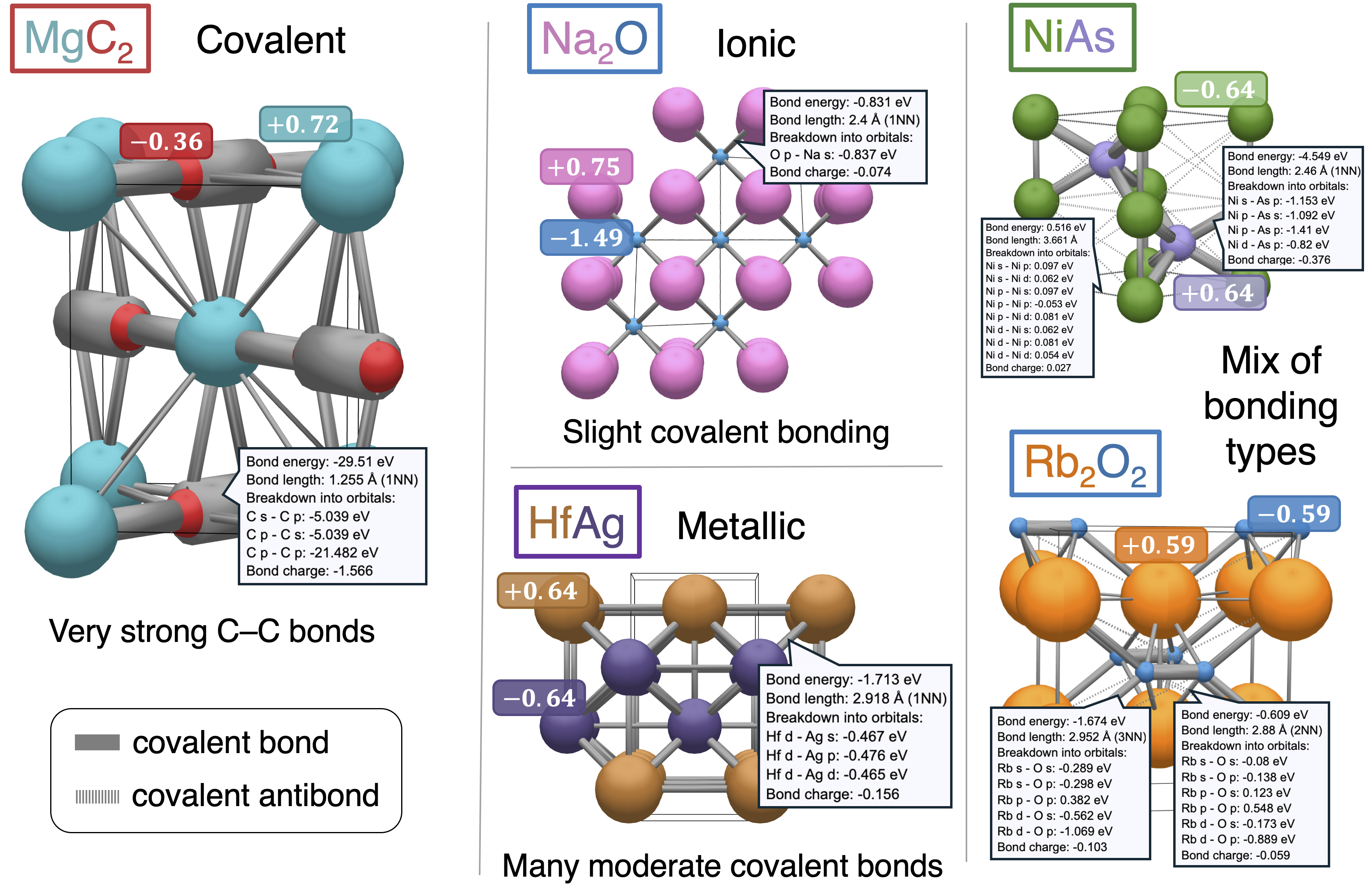}
\caption{
Crystal visualizations of the COGITO-calculated
covalent bonding for five example compounds. The crystal plots are made with bond cylinders whose radius is proportional to the COGITO iCOHP for the bond. 
Antibonds are plotted as dashed lines. Compounds are selected which span covalent, ionic, and metallic bonding regimes. Mulliken charges are displayed.}
\label{fig:vis_five}
\end{figure*}

In \textbf{Fig.~\ref{fig:vis_five}}, COGITO’s interpretable crystal bond diagrams demonstrate complex covalent interactions in compounds across different bonding regimes. Covalent bonding manifests most strongly in the carbides, represented here by MgC\textsubscript{2}, which show very strong short-range covalent bonding. Ionic bonding arises predominantly between an alkali or alkaline-earth metal and a nonmetal, represented by Na\textsubscript{2}O, where covalent bonding is very weak while charge transfer is large. Some compounds in this regime are even slightly covalently antibonding, being stabilized only by electrostatics. Metallic bonding appears when both elements are metals, represented by HfAg, where close packing and high coordination produce many moderate covalent bonds. Finally, NiAs and Rb\textsubscript{2}O\textsubscript{2} represent compounds that exhibit mixed bonding, combining short-range covalent bonds with long-range antibonding.

As exemplified by Rb\textsubscript{2}O\textsubscript{2}, the covalent bonding revealed is not always predictable from bond length. In this structure, each oxygen of the peroxy group is bonded to four Rb atoms that are 2.88 Å away and two Rb atoms that are 2.95 Å away. One would assume that the shorter bonds are stronger, but COGITO iCOHP reveals that the shorter bonds are substantially weaker, at -0.61 eV, compared to the longer bonds, at -1.67 eV. This is likely a consequence of the orbitals combining to optimize the O--O bond over the shorter O--Rb bonds, while the lone pair of the O atom opposite to the O--O bond is more readily available to form bonds with the farther Rb atoms.  

\begin{figure}[hbt!]
\centering
\includegraphics[width=6.3in]{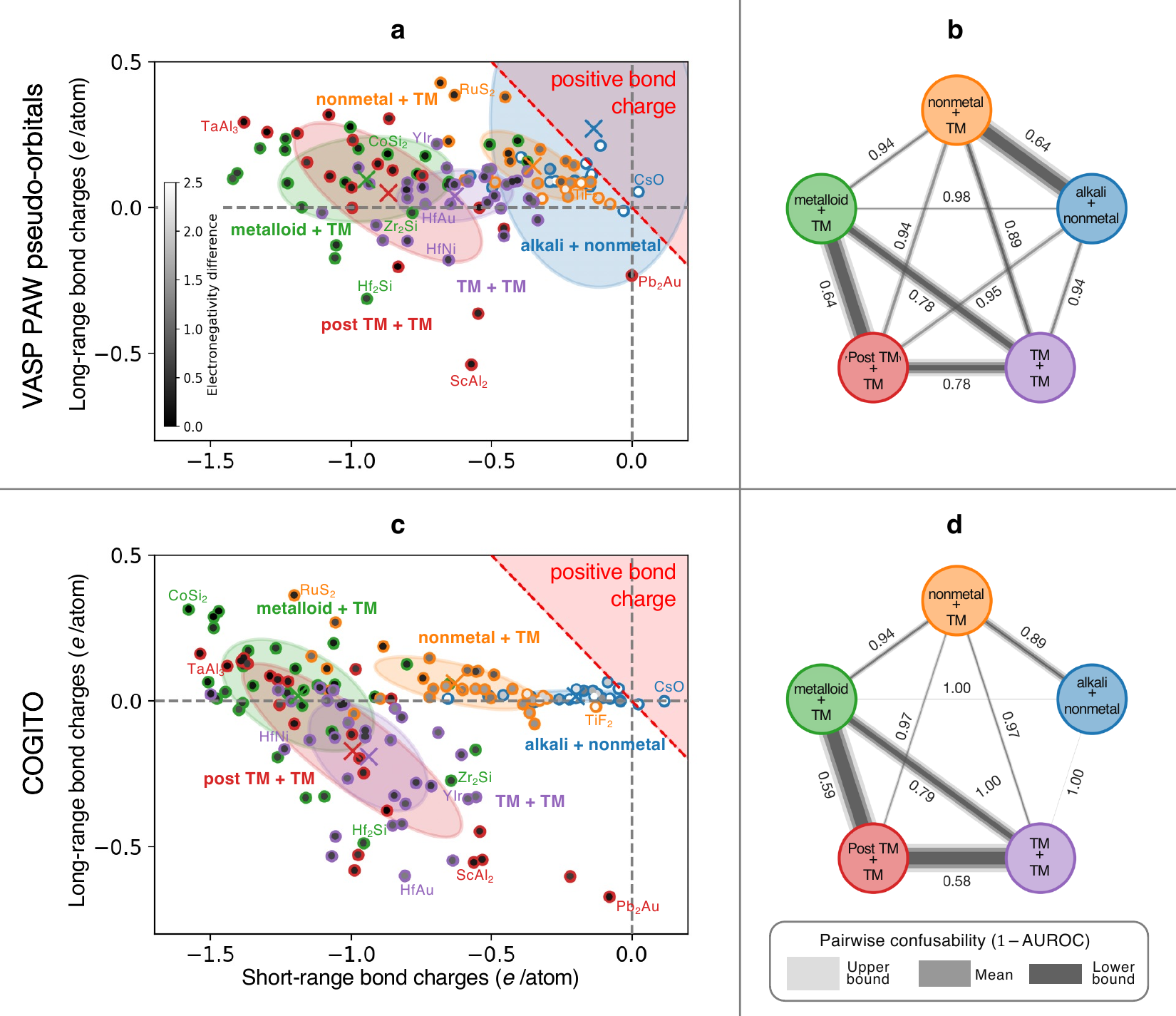}
\caption{Short-range versus long-range bond charge of binary compounds calculated with VASP PAW pseudo-orbitals (\textbf{a} and \textbf{b}) or COGITO (\textbf{c} and \textbf{d}), and the resulting group separability. 
(\textbf{a}/\textbf{c}) Integrated COOP summed over short-range and long-range bonds for each compound. Short-range is determined as $<$3 Å, or at $<$1.1× the shortest bond length when that exceeds 3 Å. 
Bonding occurs from shared electrons (negative charge), thus the lower-left quadrant corresponds to bonding while the upper-right corresponds to antibonding.
Compounds are grouped by the element class of their two constituents: alkali or alkaline-earth+nonmetal (blue), nonmetal+transition (orange), metalloid+transition (green), post-transition+transition (red), and transition+transition (purple). For sparsity, alkaline-earth is withheld from the label name and transition metal is marked as TM. 
Each marker is shaded by the electronegativity difference between the two elements. 
For each group, the ‘×’ marks the mean and the ellipse the one-standard-deviation contour of the fitted bivariate Gaussian to the groups’ compounds. 
(\textbf{b}/\textbf{d}) Pairwise confusability of the five groups, quantified by AUROC from quadratic discriminant analysis with five-fold cross-validation and equal class priors. 
Edge width encodes $1-$AUROC, so thicker edges indicate greater overlap; the three overlaid widths give the bootstrap 5th percentile, mean, and 95th percentile over 2,000 resamples of the held-out predictions. 
Compared to VASP PAW pseudo-orbitals, COGITO bond charges show increased stability,
distinguish the ionic (alkali+nonmetal) and mixed ionic-covalent (nonmetal+TM) groups, and merge the metallic groups (metalloid+TM, post TM+TM, and TM+TM).
}
\label{fig:range_bonds}
\end{figure}

To assess the reliability of covalent bond descriptors, we test how well such descriptors cluster binary compounds based on their composite elements when constructed with VASP PAW pseudo-orbitals versus COGITO. The VASP PAW pseudo-orbitals are among the most accurate predetermined bases available, frequently achieving low charge spilling and small band-interpolation errors.\cite{Oliphant_COGITO_2026} Thus, comparing COGITO against them provides a clear ablation for the added value from an \emph{adaptable} basis rather than merely an accurate one. Based on the observed bonding variation in \textbf{Fig.~\ref{fig:vis_five}}, we reduce the high-dimensional bonding data (Methods~\ref{sec:cohp_form}) into descriptors of total short-range and long-range bond charge per atom, obtained by summing the iCOOP over bonds shorter and longer than the range cutoff defined in the caption of \textbf{Fig.~\ref{fig:range_bonds}a}. Restricting the dataset to binaries formed from element-class pairs with sufficient representation leaves 111 compounds in five groups: alkali or alkaline-earth+nonmetal (A+NM, 18), nonmetal+transition (NM+TM, 22), metalloid+transition (M+TM, 22), post-transition+transition (pTM+TM, 18), and transition+transition (TM+TM, 31). 

With COGITO, the individual centers move toward greater bond charge in both coordinates (toward the lower left), shifting the average short-range (long-range) bond charge of all compounds from $-0.6e$ ($0.17e$) with PAW to $-0.8e$ ($-0.12e$) with COGITO. Most notably, A+NM shifts from substantial long-range antibonding ($0.27e$) with PAW—a center pulled further by a few outliers—to only slightly long-range antibonding ($0.02e$) with COGITO and a uniquely elongated distribution along the short-range axis. This is consistent with the A+NM compounds being strongly ionic, with minimal short-range covalency and no long-range covalency. For many metal+metal groups, COGITO not only reduces long-range antibonding but achieves substantial negative long-range bond charge, a direct signature of the delocalized electron sharing that defines metallicity. 

To quantify the separability of each group, we perform quadratic discriminant analysis (leveraging the bivariate Gaussians visualized in \textbf{Fig.~\ref{fig:range_bonds}a\&c}) with equal class priors and five-fold cross-validation. Classifying into one of the five groups, COGITO reaches 54\% accuracy compared to 44\% for PAW. However, rather than perfect five-way classification, we should expect that separability between groups mirrors their chemical relationships. Thus, we also evaluate the separability between each pair of groups using area under the receiver operating characteristic curve (AUROC). AUROC provides the probability that a randomly chosen member of group A is assigned a higher probability of belonging to A than a randomly chosen member of group B, so that 1.0 indicates complete separability and 0.5 indicates chance. The resulting pairwise confusability (1-AUROC) is shown in the connectivity graphs of \textbf{Fig.~\ref{fig:range_bonds}b\&d}, where thicker edges denote greater overlap.

Averaged over all ten pairs, separability is nearly identical for the two bases (0.85 for PAW versus 0.87 for COGITO); what changes is \emph{which} pairs are separable. The most striking difference is that COGITO renders A+NM essentially distinguishable from every other group, raising its mean pairwise AUROC from 0.86 to 0.97, and COGITO similarly sharpens NM+TM from 0.85 to 0.94. Conversely, separability between most metal+metal groups decreases with COGITO, in particular the AUROC between pTM+TM and TM+TM compounds decreases from 0.78 to 0.58. Thus, COGITO isolates the two nonmetal-containing bonding regimes while producing greater overlap among the sampled metal--metal compounds in this short-range versus long-range bond descriptor space.

\subsection{Covalency and ionicity in GaN polymorphs}
\label{sec:bonds_gan}

Next, we use COGITO to quantify the covalency versus ionicity in four structural polymorphs of GaN and observe that COGITO has an improved ability to capture chemical expectation. 
The covalency is measured using the iCOHP for 1NN
interactions (and $>$NN interactions), and the ionicity is measured using the Madelung energy with atomic charges determined with our COGITO basis and Mulliken partitioning\cite{mulliken_electronic_1955} defined in \textbf{Eqn. 13} in Methods~\ref{sec:cohp_form}. 
Our results with COGITO are compared with the LOBSTER results calculated by George et al.
\cite{george_automated_2022}.

To start, we summarize the anticipated chemistry of the four structural polymorphs: wurtzite, sphalerite, hexagonal planar (HP), and halite. We would anticipate hexagonal planar to be a highly covalent structure, given its structural similarity to graphite/graphene, and because the low 3-fold coordination is a hallmark of high covalency. 
In contrast, halite is a highly ionic structure, typified by NaCl and other strong cation-anion compositions. 
From a coordination-perspective, halite has a non-directional dense packing of ions that supports the isotropic nature of the ionic electrostatic interaction. 
Wurtzite and sphalerite should have similar bonding motifs between the extremes of HP and halite. Structural maps support this arrangement, identifying that the stable structure of a 1:1 binary compound switches from sphalerite, to wurtzite, to halite as electronegativity difference between the elements increases.\cite{Mooser_1959}

\begin{table*}[hbt]
\caption{\label{tab:GaN-polymorphs}Comparison of integrated COHP and Madelung energy for the GaN polymorphs: wurtzite, sphalerite, hexagonal planar (HP), and halite when calculated from COGITO and LOBSTER~\cite{george_automated_2022}. COGITO aligns with chemical intuition by showing a gradual increase in covalency from wurtzite to sphalerite to HP polymorphs while the halite polymorph is strikingly least covalent and \textbf{most} ionic. LOBSTER contradicts this intuition, predicting that HP is least covalent and \textbf{most} ionic while halite is least ionic.}
\begin{ruledtabular}
\begin{tabular}{lccccccc}
 &  & \multicolumn{4}{c}{COGITO} & \multicolumn{2}{c}{LOBSTER~\cite{george_automated_2022}} \\
\cline{3-6}\cline{7-8}
Structure &
\begin{tabular}[c]{@{}c@{}}Total energy\\ /Ga (eV)\end{tabular} &
\begin{tabular}[c]{@{}c@{}}1NN iCOHP\\ /Ga (eV)\end{tabular} &
\begin{tabular}[c]{@{}c@{}}$> $NN iCOHP\\ /Ga (eV)\end{tabular} &
\begin{tabular}[c]{@{}c@{}}Charge \\ ($e$)\end{tabular} &
\begin{tabular}[c]{@{}c@{}}Madelung\\/GaN (eV)\end{tabular} &
\begin{tabular}[c]{@{}c@{}}Ga--N iCOHP\\ /Ga (eV)\end{tabular} &
\begin{tabular}[c]{@{}c@{}}Madelung\\ /GaN (eV)\end{tabular} \\
\hline
Wurtzite & \textbf{-12.164}  & -17.22 & 2.00 & 0.821 & -8.08 & -20.12 & -11.61 \\
Sphalerite & -12.154   & -17.37   & 2.03 & 0.783 & -7.35 & \textbf{-20.20} & -10.85 \\
HP      & -11.455    & \textbf{-18.02}   & \textbf{1.50} & 0.817 & -7.99 & -18.75 & \textbf{-15.11} \\
Halite      & -11.206  & -14.93  & 3.28 & \textbf{0.905} & \textbf{-9.66} & -19.23 & -9.16 \\
\end{tabular}
\end{ruledtabular}
\label{tab:cog_lob}
\end{table*}

\begin{figure*}[hbt!]
\centering
\includegraphics[width=6.4in]{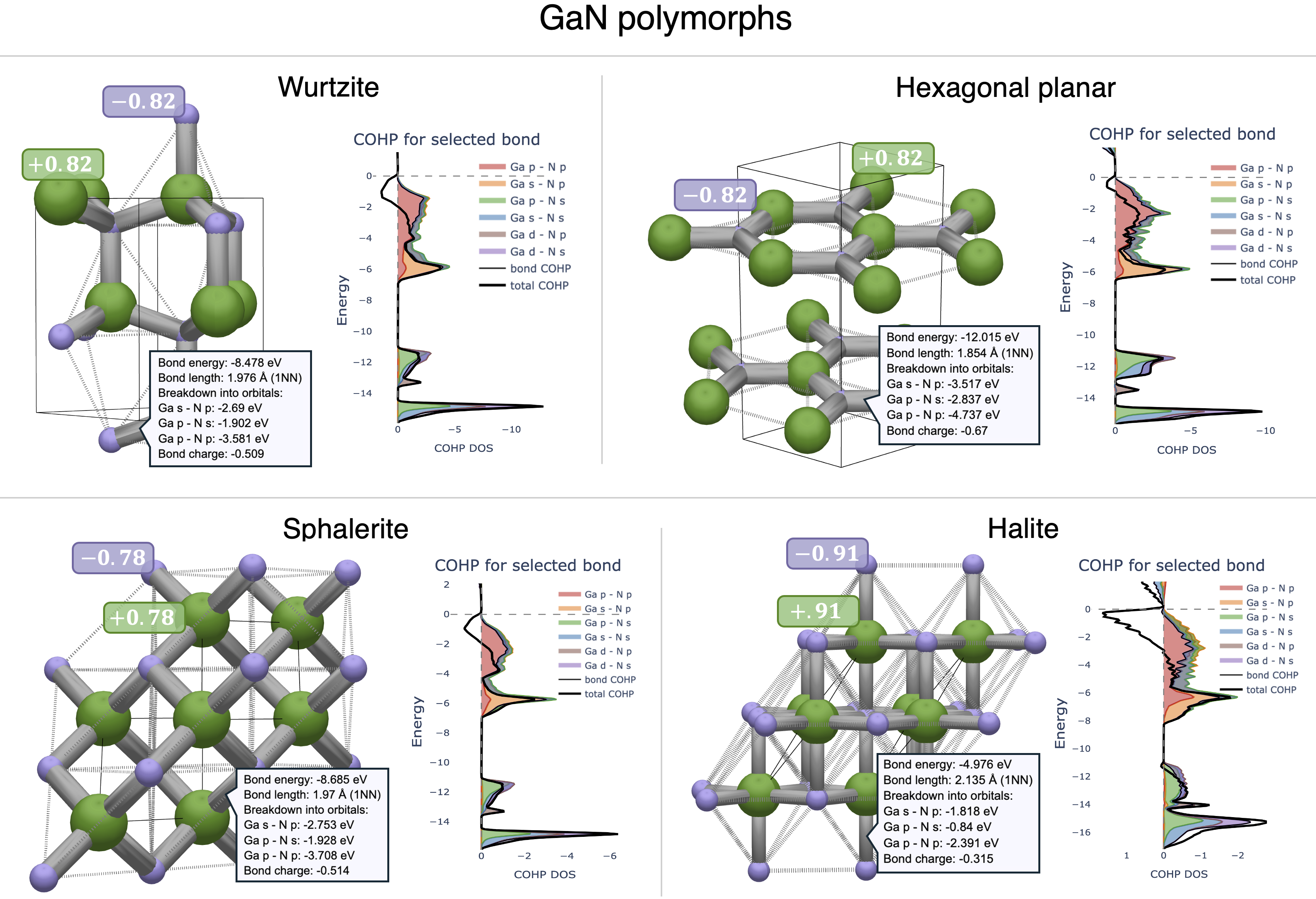}
\caption{Crystal visualizations for the COGITO-calculated
covalent bonding and charge transfer in GaN polymorphs.
The COHP-projected density of states is plotted for all interactions (total COHP) and just the 1NN Ga--N bonds (bond COHP). 
The bond COHP is decomposed into individual orbital-orbital COHP stacked on top of each other to sum to the full bond COHP. From the crystal bond plot, we see that wurtzite, HP, and sphalerite have similar total 1NN bonds and charge transfer, while halite is substantially less covalent and more ionic. 
The difference between the bond COHP DOS and the total COHP DOS reveals that halite has substantial 2NN+ antibonding while HP has minimal 2NN+ antibonding.}
\label{fig:gan}
\end{figure*}

As summarized in the right side of \textbf{Table~\ref{tab:cog_lob}}, George et al. used LOBSTER's Ga--N iCOHP to measure covalency, and the Madelung energy calculated from LOBSTER-derived Mulliken charges to measure ionicity. 
Of the four polymorphs, LOBSTER's results indicate that the hexagonal planar structure is the least covalent and the most ionic, that the halite structure is the least ionic, and that the sphalerite structure is the most covalent.
LOBSTER's results for HP and halite strongly contradict our established expectation. 
While LOBSTER captures the similarity between wurtzite and
sphalerite, it places them incorrectly relative to the other structures, with
HP being, on average, 1.4 eV less covalently stable while halite is, on average, 2.1 eV less
electrostatically stable. 
Despite this, LOBSTER's emphasis on usability and flexibility has enabled countless impactful studies of chemical bonding in
materials.\cite{zheng_essential_2024,zhao_plasticity_2024,zhao_structural_2020,zhan_highly_2024,guo_tackling_2020,han_engineering_2021}

COGITO's analysis of iCOHP (covalency) and charge transfer (ionicity) for wurtzite, sphalerite, HP, and halite is shown in \textbf{Fig.~\ref{fig:gan}} and summarized in \textbf{Table~\ref{tab:cog_lob}}. To match the convention of LOBSTER, we take half of the total iCOHP values. This accounts for only one of the symmetric bond pairs $AB\mathbf{R}$ and $AB\overline{\mathbf{R}}$, where $\overline{\mathbf{R}}$ indicates $-\mathbf{R}$.
The wurtzite and sphalerite structures have comparable covalent bond strengths, with sphalerite being 0.15 eV more stable for 1NN bonding while wurtzite is 0.03 eV more stable for $>$NN bonding. Consistent with LOBSTER and previous expectations \cite{Mooser_1959,Yeh_zbwurt_1992}, we find wurtzite to be more electrostatically stable than sphalerite, with 0.038 e$^{-}$ more charge transfer leading to a Madelung energy that is 0.73 eV more stable. 

COGITO identifies HP as the most covalent structure, with a total iCOHP that is 1.30 eV more stable than wurtzite (0.80 eV from 1NN with an additional 0.50 eV from $>$NN). 
Despite its high covalency, HP still boasts a charge transfer just less than wurtzite. 
This comparable charge transfer could be explained by HP's ability to stabilize certain ionic compounds (e.g. BN). However, the apparent stability of HP relative to wurtzite (judged by the simple sum of the 1NN iCOHP and Madelung energy) is something to be cautious of. 
Fortunately, COGITO does reduce this apparent stability of HP (0.71 eV more stable than wurtzite, versus 2.13 eV with LOBSTER). 
The halite structure stands out from the other GaN polymorphs as the
most ionic, with 0.905 electrons transferred, and the least
covalent, with its total covalent bonding being 3.57 eV/Ga less stable than wurtzite. 
Overall, COGITO successfully captures the key chemical
intuition of these GaN polymorphs, while providing additional bonding
insights apparent from COGITO's robust and intuitive visualizations of
its high-dimension COHP data.

\newpage
To isolate the cause of differences between LOBSTER and COGITO, \textbf{Table~\ref{tab:paw}} shows the same iCOHP and charge data acquired through the COGITO package but when projecting the VASP PAW pseudo-orbital without any iteration. 
Changes between the LOBSTER and PAW values result from differences in the selected basis and technical implementation, including LOBSTER's orthogonalization before COHP analyses (as of LOBSTER version 2.0),\cite{maintz_lobster_2016,maintz_analytic_2013} and construction of
\textbf{k}-dependent Hamiltonians by projection onto all-electron
orbitals\cite{maintz_analytic_2013} rather than PAW transformed pseudo-orbitals\cite{agapito_accurate_2016}. 
Whereas, changes between PAW and COGITO are purely a result of COGITO using an atomic basis which adapts to local differences between the polymorphs. 
Although PAW takes a step towards chemical expectation compared to LOBSTER (the Madelung energy of HP becomes less stable while halite becomes more stable and the 1NN iCOHP of halite becomes less stable), it is not enough to obtain the correct qualitative trends. In addition, the PAW pseudo-orbitals produce, on average, 5× more antibonding for $>$NN iCOHP compared to COGITO, considerably over-predicting long-range destabilizing interactions which the COGITO basis reduces.

\begin{table}[!b]
\centering
\setlength{\tabcolsep}{3pt}
\caption{\label{tab:GaN-PAW}
Integrated COHP, charge transfer, and Madelung energy for the GaN polymorphs: wurtzite, sphalerite, hexagonal planar (HP), and halite when calculated with VASP PAW pseudo-orbitals.
}
\begin{tabularx}{0.5\columnwidth}{lcccc}
\hline\hline
 &  \multicolumn{4}{c}{PAW} \\
\cline{2-5}
Structure &
\begin{tabular}[c]{@{}c@{}}1NN iCOHP\\ /Ga (eV)\end{tabular} &
\begin{tabular}[c]{@{}c@{}}$> $NN iCOHP\\ /Ga (eV)\end{tabular} &
\begin{tabular}[c]{@{}c@{}}Charge \\ ($e$)\end{tabular} &
\begin{tabular}[c]{@{}c@{}}Madelung\\/GaN (eV)\end{tabular}  \\
\hline
Wurtzite   & -18.69 & 11.28 & 0.945 & -10.72 \\
Sphalerite & \textbf{-18.75} & 11.37 & 0.925 & -10.25 \\
HP        & -17.77 & \textbf{7.35} & \textbf{1.045} & \textbf{-13.08}\\
Halite    & -16.05 & 13.07 & 0.940 & -10.41 \\
\hline\hline
\end{tabularx}
\label{tab:paw}
\end{table}

\subsection{Range-resolved origins of the silicon band
structure}
\label{sec:bonds_si}

\begin{figure*}[hbt!]
\centering
\includegraphics[width=7in]{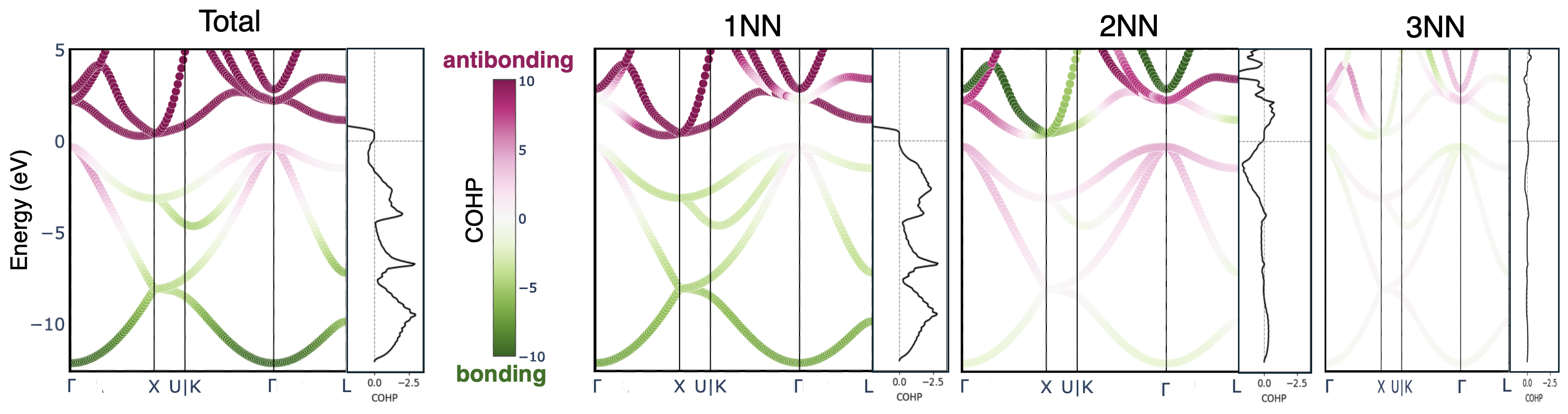}
\caption{The COHP-projected band structure and density of
states for silicon. The COHP is separated into first, second, and third
nearest neighbors, showing how different neighboring arrangements
contribute to the shape of the bands.}
\label{fig:cohp_si}
\end{figure*}

The silicon band structure has been examined many times from a
tight-binding approach.\cite{chadi_tight-binding_1975,ciraci_electronic-energy-structure_1977,grosso_tight-binding_1995,lenosky_highly_1997,sapra_realistic_2002,green_silicon_2005,niquet_onsite_2009} However, most previous attempts
have focused only on first nearest-neighbor interactions. Recently, we
demonstrated that the indirect bandgap of silicon depends intimately on
second nearest-neighbor interactions.\cite{oliphant_why_2025} Here, for the first time,
we show the COHP-projected band structure for bonds with the first nearest-neighbor
(1NN), second nearest-neighbor (2NN), and third nearest-neighbor (3NN) atoms in \textbf{Fig.~\ref{fig:cohp_si}}. In other
words, COGITO enables us to identify the bonding and antibonding
character of individual bands in the band structure, as a function of
nearest-neighbor interaction. Isolating the 2NN interactions is not possible when
only using projected \textbf{k}-dependent Hamiltonians. 
Since the 2NN interactions are between a silicon atom and its translation in neighboring cells, the 2NNs get mapped to the onsite term of the Bloch orbital interacting with itself. 
Similarly, the 3NN interactions get mapped onto the 1NN terms when only projection is used.

Using COGITO's ability to separate all atomic interactions, we explore
how the valence band maximum (VBM) and conduction band minimum (CBM)
form as a result of the crystal chemistry. This insight into the effect
of bonds on band extrema can be used to manipulate the band gap by
engineering the chemistry or short-range coordination
chemistry.\cite{franceschetti_inverse_1999} One may assume that the VBM \emph{p}-orbital
wavefunction is bonding since it is by definition more stable than its
conduction band counterpart at $\Gamma$. However, while COGITO shows 1NN
bonding for most of the valence bands, it reveals that the VBM is
slightly antibonding for 1NN while the above conduction band
wavefunction is bonding. This oddity is explained by examining the 2NN
COHP, which shows that the bonding arrangement for 1NN (lowest
conduction band at $\Gamma$) leads to significantly more antibonding in the 2NN
interaction, thus stabilizing the 1NN antibonding arrangement relative
to the 1NN bonding arrangement. Additionally, the 3NN interaction is
bonding (at VBM) in the 1NN antibonding arrangement.

The position of the silicon CBM being $\sim$0.85 along the $\Gamma$--X
line in silicon arises from a complex combination of orbital
interactions. Historically, simple 1NN tight-binding models were only able to
reproduce an indirect CBM by adding excited orbitals like 4\emph{s*} or
3\emph{d} orbitals.\cite{vogl_semi-empirical_1983,jancu_empirical_1998,boykin_valence_2004,soccodato_machine_2024} Recently, we demonstrated an approach
to disentangle from a DFT-derived tight-binding model what important orbital
interactions contribute to band energy, effectively sifting through
hundreds of interactions.\cite{oliphant_why_2025} Using a MLWF-derived tight-binding model, we
identified that the CBM is explained by the 2NN
\emph{p\textsubscript{x}-p\textsubscript{x}} interactions pulling down
the band near the X point. However, our analysis required careful
checking that the orthogonal Wannier function model aligned with
expected values from the original projected atomic orbitals. In some
cases, they did not align. For example, we found that the MLWF 2NN
\emph{s--s} hopping parameter was destabilizing, something which is
inconsistent with \emph{s} orbitals since they contain no angular nodes.
The 1NN \emph{p\textsubscript{x}-p\textsubscript{x}} and 2NN \emph{s-- p\textsubscript{x}} Hamiltonian parameters also switch
signs between COGITO and MLWF. 
This alteration in the perceived bond type between \emph{s}-like or \emph{p}-like Wannier
functions is a result of the Wannier functions' oscillating tails around
neighboring atoms.\cite{Oliphant_COGITO_2026} 
Additionally, COGITO produces larger COHP magnitudes for antibonding states than MLWF (or LOBSTER) because the nonorthogonal orbital coefficients become larger to compensate for the loss in electron normalization from antibonding.

Now using the COGITO tight-binding model to reliably represent local atomic
orbitals and our implementation of projected COHP, we can verify the
origin of the CBM in silicon. Examining the lowest conduction band along
$\Gamma$--X in \textbf{Fig.~\ref{fig:cohp_si}} shows that the 1NN interactions have near zero
effect at $\Gamma$ but are substantially antibonding at X, confirming that the
1NNs destabilize the energy at X. Next, we observe that the 2NNs are
antibonding at $\Gamma$ and gradually switch to bonding at X, confirming that
the 2NN are the major stabilizing contribution to the indirect band gap
in silicon. Of the 2NNs, COGITO finds only the
\emph{p\textsubscript{x}-p\textsubscript{x}} interaction lowers the band
at X compared to $\Gamma$. Our analysis with COGITO also reveals that the 3NNs
further bring down the CBM near X, even decreasing the energy most at
$\sim$85\% of the way to X, although it is 2.5\(\times\) weaker
than the 2NN contribution. With COGITO, this clear breakdown of band structure into range-resolved bonding contribution can be done with any material without the tedious chemical consistency checks that were necessary with orthogonal Wannier methods.

\section*{Discussion}

COGITO (Crystal Orbital Guided Iteration To atomic-Orbitals) converts the periodic Kohn–Sham electronic structure into chemically interpretable quantities. 
The essential distinction from Wannier approaches or conventional projection is the simultaneous preservation of a basis that remains strictly atomic, radial functions and overlaps that adapt to the crystal environment, and Hamiltonian, overlap and density matrices resolved over lattice translations. 
Building on earlier COGITO work, we improved and validated the fitting procedure so that numerical artifacts — nodal tails of Wannier orbitals, integration-grid choice, supercell size — are suppressed without damping the physically meaningful orbital relaxation that distinguishes an adaptive basis from a fixed one.

The advantage of COGITO’s adaptable atomic solution is benchmarked against VASP PAW pseudo-orbitals, LOBSTER, and Wannier90/MLWF. The clearest contrast appears at long range (3–5 Å). 
PAW pseudo-orbitals generate spurious long-range antibonding. By comparison, COGITO suppresses long-range antibonding in covalent and ionic compounds and identifies more long-range bonding for metal–metal systems, consistent with delocalized metallic bonding. 
Orthogonalized Wannier schemes are further removed from chemical reliability, returning long-range Hamiltonian elements whose magnitude and sign are inconsistent with atomic-orbital expectations. 
Conversely, because COGITO has full atomic reliability, it enables band dispersion to be decomposed into distance-resolved contributions without case-by-case validation. 
With this, we confirm that the silicon conduction-band minimum is lowered predominantly by the second-nearest neighbors and identify weaker third-neighbor contributions that further lower the low-symmetry minimum.

COGITO descriptors also resolve ionicity, covalency, and metallicity more sharply, whether changes in bonding result from variations in atomic composition or crystal structure. 
In supervised classification of compound groups based on COGITO short- and long-range bond charge, the highly ionic group (alkali–nonmetal) and the mixed ionic-covalent group (nonmetal–transition metal) separate cleanly (average pairwise AUROC of 0.97 and 0.94). 
Meanwhile, predominantly metallic groups (which are entirely metalloid, post transition metal, and transition metal atoms) remain entangled (0.58–0.79). 
For the GaN polymorphs, LOBSTER and PAW pseudo-orbitals invert the covalency–ionicity ordering which COGITO recovers.

The COGITO framework produces explicit bonding rather than latent networks. 
However, these descriptors remain limited by the approximations of the underlying DFT calculations and of the chosen population and bonding analyzes. 
We anticipate that descriptors built from COGITO will greatly enable physically grounded features for material informatics and highlight the orbital interactions on which a targeted property depends.

\renewcommand{\thesubsection}{\Alph{subsection}} 
\setcounter{subsection}{0} 
\newpage
\section*{Methods}

\subsection{Changes in COGITO for improved robustness}
\label{sec:cog_robust}
Two substantial changes have been added to COGITO that ensure a consistent COGITO basis for variable noise in the local environment or size of lattice unit cell. 
To validate the new COGITO basis, we show an example comparison of various functional fits to noised VASP PAW pseudo-orbitals and the sensitivity of the Wannier orbital fits for the GaN polymorphs (plus a supercell of sphalerite) to grid resolution. 

In the original COGITO manuscript, section II.3 avoided the computational cost of calculating Wannier functions by approximating local atomic orbitals from the \textbf{k}=0 Bloch atomic orbitals. 
Unfortunately, the quality of this Bloch-to-local approximation, and the portion of the Brillouin zone that is included at \textbf{k}=0, improves with the size of the unit cell, leading to different solutions for different supercell sizes. 
To resolve this, COGITO (as of version 0.3.0) performs the atomic orbital fitting directly on the Wannier orbitals. Computational and memory expense is mitigated by only constructing the Wannier orbitals on centered, local real-space grids for each of the irreducible \textbf{k}-points and transforming the weighted sum by each symmetry operation of the crystal lattice (changing atom and orbital type as needed).

Next, to limit spurious variability from noise or grid sampling, we designed a new fitting function (\textbf{Eqn. 2}) whose functional space is more tightly restricted to smoothly decaying orbitals while retaining sufficient flexibility to capture changes in orbital shape due to charge transfer or local environment. These changes are included as of 
Noise occurs in Wannier orbitals because of nodal tails that form around neighboring atoms as a result of the fixed overlap constraint detailed in Sec I of Ref. \cite{Oliphant_COGITO_2026}. 
Additional variability arises when the three-dimensional sampling grid is mapped onto the radial grid for fitting. Because the grid follows the generally nonorthogonal unit-cell vectors, its radial point distribution depends on the cell geometry and grid resolution.

\begin{align}
R_{\alpha}(r)
&= r^{l}\, [a_{\alpha}\,e^{-b_{\alpha}r^{2}} + c_{\alpha}\,e^{-d_{\alpha}r^{\gamma}}\tanh(\frac{r}{f})^{2}].
\end{align}

This form is selected because it permits a more gradual decay ($\gamma\rightarrow1$) in the orbital while still ensuring that the derivative of the radial function (without $r^{l}$) at $r=0$ is zero, which is consistent with the VASP PAW pseudo-orbitals. Additional constraints on the parameters ensure that both parts contribute similarly to the radial function magnitude, that both parts contribute similarly to the curvature of the radial function at $r=0$, and that the second, gradual decay part ``turns on'' before decaying. The new, constrained radial function has been robustly incorporated into COGITO as of version 0.3.13.

\begin{figure}[b!]
\centering
\includegraphics[width=7.0in]{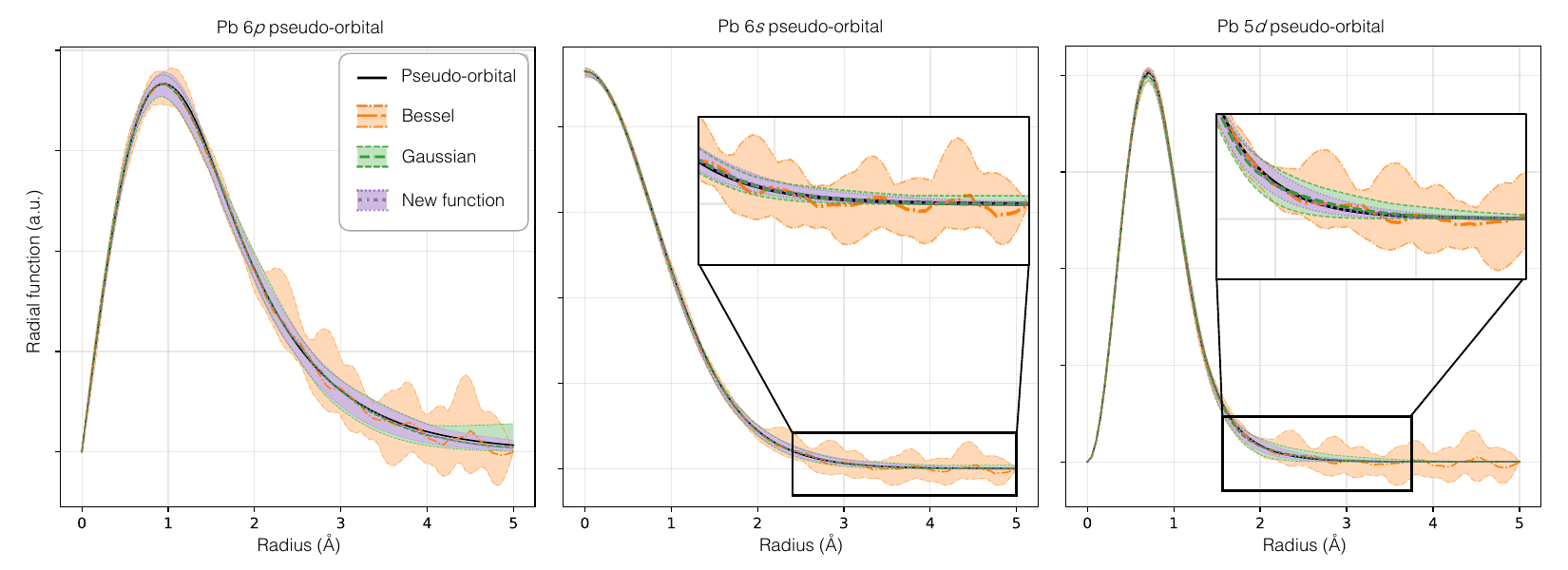}
\caption{Fits of 16 Bessel functions (orange), three Gaussian functions (green), and our new function (purple) to noised VASP PAW 6\textit{p}, 6\textit{s}, and 5\textit{d} pseudo-orbitals (black) for the Pb atom. The shaded region shows the widest envelope of solutions over 25 different noise seeds. The central line shows the average.}
\label{fig:radial_fits}
\end{figure}


We test our new alternative function against two common options: a sum of three Gaussians and a sum of 16 orthogonal spherical Bessel functions. Similar to our previous manuscript, the sum of Gaussians is partially ordered, with one gaussian being allowed to be negative with a constraint to ensure that the total sum does not have a radial node. The fit with Bessel functions is a useful comparison because their orthogonality makes their least squares solution exact, but this unique solution may still be perturbed, perhaps substantially, by noise in the fit data.

As an illustrative test, we fit each function to the Pb 6\textit{p}, 6\textit{s}, and 5\textit{d} VASP PAW pseudo-orbital noised by $0.01\sqrt{r}$. The increase of noise with radius is to mimic the noise in Wannier orbitals which is larger at higher $r$ due to mixing with the surrounding atomic orbitals. The VASP PAW pseudo-orbitals are normalized to one (without the PAW correction) for a consistent application of the noise. 
In \textbf{Figure~\ref{fig:radial_fits}} and \textbf{Table~\ref{tab:noise}}, we show the variability of the radial function and the fit parameters when the noise is generated from 25 different seeds. 
The fit of the 16 Bessel functions is highly sensitive to noise, creating substantial oscillations towards the tail end of the radial function. 
While both the Gaussian and new function reduce oscillations, the Gaussian fit has a wider spread than the new function and at times finds a fit which decays very slowly at long-range.

\begin{table}[bt!]
\centering
\begin{tabular}{ccccc}
\hline
\hline
Method & Orbital & Min RSD & Max RSD & Mean RSD \\
\hline
\hline

{Bessel}
  & \textit{s} & 0.076 & 1.351 & 0.891 \\
  & \textit{p} & 0.073 & 1.447 & 0.981 \\
  & \textit{d} & 0.095 & 1.342 & 0.621 \\
\cline{1-5}

{Gaussian}
  & \textit{s} & 0.111 & 1.621 & 0.691 \\
  & \textit{p} & 0.183 & 1.187 & 0.564 \\
  & \textit{d} & 0.030 & 1.236 & 0.428 \\
\cline{1-5}

{New function}
  & \textit{s} & 0.035 & 0.184 & 0.104 \\
  & \textit{p} & 0.050 & 0.218 & 0.126 \\
  & \textit{d} & 0.025 & 0.144 & 0.085 \\
\hline
\hline
\end{tabular}
\caption{The minimum, maximum, and mean relative standard deviation (RSD) of model parameters over the 25 random seeds.}
\label{tab:noise}
\end{table}

\begin{figure}[b!]
\centering
\includegraphics[width=4.0in]{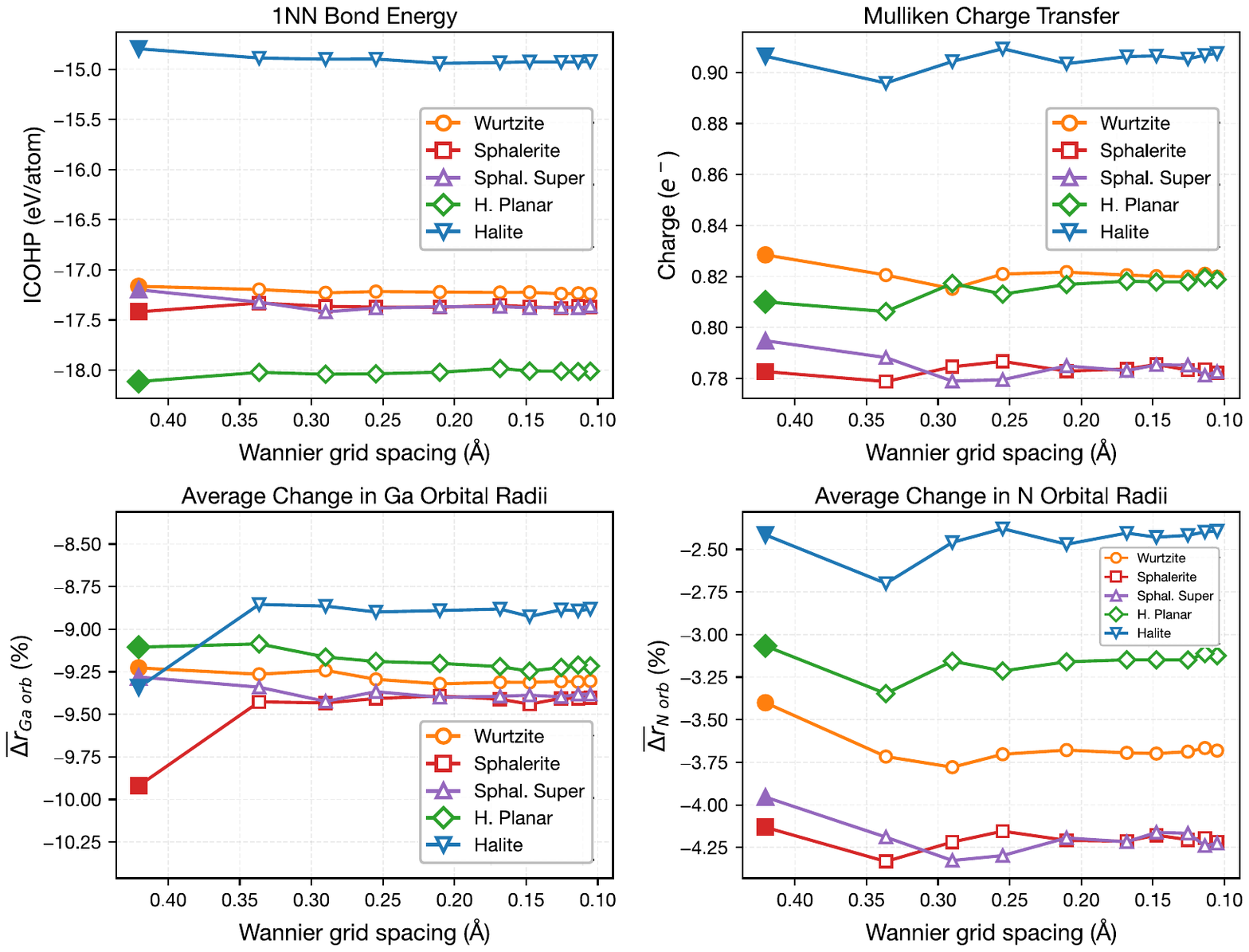}
\caption{Convergence of COGITO 1NN bond energy, Mulliken charge transfer, average Ga orbital radii, and average N orbital radii with Wannier grid spacing for four GaN polymorphs: wurtzite, sphalerite, hexagonal planar, and halite. Sphalerite is also run in a supercell to examine dependence on cell size and shape. The metrics achieve good convergence by around 0.25 Å spacing, with differences between polymorphs being over an order of magnitude larger than variation from further decrease in grid spacing.}
\label{fig:gan_conv}
\end{figure}

In \textbf{Table~\ref{tab:noise}}, we summarize how the 25 different noise seeds perturb the parameters of each model (e.g. $a$, $b$, $c$, $d$, $\gamma$, and $f$ in Eqn. 1) by calculating the relative standard deviation (RSD) for each parameter across the seeds and reporting the minimum, maximum, and mean RSD for the Pb 6\textit{s}, 6\textit{p}, and 5\textit{d} orbital parameters.
Despite their unique solution per noise seed, the Bessel function parameters experience a large perturbation across different noise seeds.
The RSD is generally low ($<0.1$) for the parameters of Bessel functions with few nodes but high ($>1.0$) for the parameters of Bessel functions with many nodes. 
As a whole, this leaves the fit with Bessel functions highly sensitive to noise, with a mean RSD across all orbitals of 0.831. 
The Gaussian model partly improves, with a lower mean RSD across all orbitals of 0.561, but the maximum RSD remains above 1.0. 
In contrast, the new function yields the lowest minimum, maximum, and mean parameter RSDs, with an overall mean of 0.105.
This greater parameter stability supports interpreting changes in the fitted orbital basis across crystal structures with reduced interference from numerical noise.

Finally, to verify the stability of the new function, \textbf{Figure~\ref{fig:gan_conv}} shows the convergence of bond energy, charge transfer, and orbital radii with decreasing Wannier grid spacing for four GaN polymorphs: wurtzite, sphalerite, hexagonal planar, and halite. 
To test variation with cell size and shape, we include COGITO calculations on a four-atom supercell of sphalerite which has a third cell vector that is orthogonal to the first and second cell vectors, as in the wurtzite primitive cell. 
In comparison, the two-atom primitive cell of sphalerite has three mutually nonorthogonal cell vectors, as in the diamond primitive cell.
Overall, the metrics achieve good convergence by around 0.25 Å spacing. 
Although slight variations in the metrics between the grid spacings sampled remain, differences between polymorphs are over an order of magnitude larger.  

\subsection{Building a local Hamiltonian from KS-DFT}
\label{sec:local_ham}

Next, we now briefly review how to transform the KS wavefunctions into an atomic orbital frame, and create a local orbital Hamiltonian from said transformation. 

Before operating locally, we must first work within the \textbf{k}-space of periodic systems. This requires that we extend local atomic orbitals to Bloch atomic orbitals, which are the sum of local atomic orbitals in all translated unit cells \textbf{R} multiplied by a phase factor determined by \textbf{k}-point.

\begin{equation}
\Phi_{\alpha}^{\mathbf{k}}
= \sum_{\mathbf{R}} e^{i\mathbf{k}\cdot(\mathbf{R}+\boldsymbol{\tau}_{\alpha})}\,
\phi_{\alpha}^{\mathbf{R}+\boldsymbol{\tau}_{\alpha}}.
\tag{3}
\end{equation}
where \(\phi_{\alpha}^{\mathbf{R}+\boldsymbol{\tau}_{\alpha}}\) is a local atomic orbital centered at \(\mathbf{R}+\boldsymbol{\tau}_{\alpha}\).

A \textbf{k}-dependent Hamiltonian of these Bloch orbitals can be constructed directly from \(\Phi_{\alpha}^{\mathbf{k}}\) if one has access to the action of the Hamiltonian operator.\cite{Gong2024} But in practice, it is useful to transform the action of the Hamiltonian operator to the basis KS wavefunctions, where it is diagonal, and project the Bloch atomic orbitals onto the predetermined KS wavefunctions. Any operator can be transformed from the KS basis into the Bloch orbital basis in an equivalent manner.

\begin{equation}
H_{\alpha\beta}^{\mathbf{k}}
\equiv \braket{\Phi_{\alpha}^{\mathbf{k}}|\widehat{H}|\Phi_{\beta}^{\mathbf{k}}}
\approx \sum_{n,m}\braket{\Phi_{\alpha}^{\mathbf{k}}|\psi_{n}}\braket{\psi_{n}|\widehat{H}|\psi_{m}}\braket{\psi_{m}|\Phi_{\beta}^{\mathbf{k}}}
= \sum_{n}\braket{\Phi_{\alpha}^{\mathbf{k}}|\psi_{n}}\,\varepsilon_{n}^{\mathbf{k}}\,\braket{\psi_{n}|\Phi_{\beta}^{\mathbf{k}}}.
\tag{4}
\end{equation}

How closely a \(H_{\alpha\beta}^{\mathbf{k}}\) calculated from the KS wavefunctions equates to a \(H_{\alpha\beta}^{\mathbf{k}}\) calculated directly from \(\Phi_{\alpha}^{\mathbf{k}}\) depends on how well the KS wavefunctions form a complete basis for the Bloch atomic orbitals. We coined this non-completeness of the KS wavefunctions as orbital mixing in Sec. I (Eqn. 7) of Ref. \cite{Oliphant_COGITO_2026} since a consequence of possible non-completeness is that it mixes the orbital character. COGITO uniquely optimizes orbital mixing by iteratively adapting the projected atomic orbital basis to the Wannier orbitals. 

However, the accuracy of the reconstructed \(H_{\alpha\beta}^{\mathbf{k}}\) in reproducing the KS valence band energies depends mostly on the charge spilling of the atomic orbital basis and on the orthogonality between low- and high-energy KS subspaces (Fig 3 of Ref. \cite{Oliphant_COGITO_2026}). Unfortunately, the ability to optimize the accuracy of band interpolation is limited when only modifying the form of \(\Phi_{\alpha}^{\mathbf{k}}\). As a solution, after COGITO finds an updated Bloch atomic orbital basis to project, we directly modify the projections before constructing \(H_{\alpha\beta}^{\mathbf{k}}\) (see details in Sec. II.2 of Ref. \cite{Oliphant_COGITO_2026}). 

Lastly, we construct our \textbf{R}-dependent local atomic orbital Hamiltonian as a discrete Fourier transform over the \textbf{k}-dependent Bloch atomic orbital Hamiltonian.

\begin{equation}
H_{\alpha\beta}^{\mathbf{R}}
= \frac{1}{N_k}\sum_{\mathbf{k}} H_{\alpha\beta}^{\mathbf{k}}\,
e^{- i\mathbf{k}\cdot(\mathbf{R}-\boldsymbol{\tau}_{\alpha}+\boldsymbol{\tau}_{\beta})}.
\tag{5}
\end{equation}
This local Hamiltonian \(H_{\alpha\beta}^{\mathbf{R}}\) represents the interaction of local atomic orbitals \(\alpha\) (always \textbf{T}=\textbf{0} cell) and \(\beta\) (in \textbf{T}=\textbf{R} cell) with the Kohn–Sham Hamiltonian. 
Interestingly, the ideal \(H_{\alpha\beta}^{\mathbf{R}}\) for reproducing KS band energies does not perfectly achieve the Hamiltonian of the original projected atomic orbital basis \(\braket{\phi_{\alpha}^{\boldsymbol{\tau}_{\alpha}}|H_{KS}|\phi_{\beta}^{\boldsymbol{\tau}_{\beta}+\mathbf{R}}}\). 
This illustrates why the Wannier transformation is so useful, it allows one to access local quantities without being constrained to predefined orbitals. 
On the other hand, if an explicitly calculated orbital Hamiltonian is used, it takes substantially more orbitals (double/triple zeta in SIESTA\cite{Soler_SIESTA_2002}) to achieve the same accuracy in interpolated band energies as Wannier orbitals.\cite{Gong2024,Oliphant_COGITO_2026}
Furthermore, using multiple sets of orbitals restricts our ability to disentangle chemical insights. 

By solving the generalized eigenvalue problem with our \textbf{k}-dependent Hamiltonian and overlap, we can get back energies \(\varepsilon_{n}^{\mathbf{k}}\) and eigenvectors \(c_{\alpha n}^{\mathbf{k}}\) which closely resemble the original KS states. Here, \(c_{\alpha n}^{\mathbf{k}}\) encodes the chemically meaningful quantity of how much of orbital \(\alpha\) is in band \(n\).
Finally, the band energies can be expanded within the COGITO basis using the
coefficients \(c_{\alpha n}^{\mathbf{k}}\) and the \textbf{k}-dependent orbital Hamiltonian
\(H_{\alpha\beta}^{\mathbf{k}}\) as:

\begin{align}
\varepsilon_{n}^{\mathbf{k}}
= \sum_{\alpha,\beta}{c_{\alpha n}^{\mathbf{k}}}^{\dagger}\,H_{\alpha\beta}^{\mathbf{k}}\ c_{\beta n}^{\mathbf{k}}.
\tag{6}
\end{align}

Lastly, let us briefly describe the necessity to account for orbital center $\boldsymbol{\tau}_{\alpha}$ in the phase of \textbf{Eqns. 3} and \textbf{5}. If the orbital phase is not included directly, it is absorbed into the coefficients for each orbital \(c_{\alpha n}^{\mathbf{k}}\). For an orthogonal basis, this change in orbital coefficient is precisely the orbital center phase such that the downstream COHP and COOP matrices are exactly the same. However, with a nonorthogonal basis, the change in coefficient is obfuscated by the phases of the \textbf{k}-dependent overlap matrix such that elements of the COHP and COOP matrices are perturbed from the version which explicitly includes orbital center phase. This transformation is covered in more detail at the end of Sec. IV.3 in Ref. \cite{Oliphant_COGITO_2026}.

\subsection{Selecting the form of COHP/COOP}
\label{sec:cohp_form}

The nine-dimensional COHP of \textbf{Eqn. 1} is understood and
visualized by integrating over select dimensions. For example, in
\textbf{Eqn. 7}, summing over KS bands and \textbf{k}-points
gives the five-dimensional integrated COHP (iCOHP) which describes the
total band energy contribution from atomic orbital \(\alpha\)
interacting with atomic orbital \(\beta\) in cell \textbf{R}. The
occupation
\(f_{n\mathbf{k}} = \{ 1\ \mathrm{if}\ \varepsilon_{nk} \leq E_{F},0\ \mathrm{if}\ \varepsilon_{nk} > E_{F}\}\)
is used to only include bands that are below the Fermi energy.
The \textbf{k}-point weight is \(w_{\mathbf{k}}\).
Alternatively, \textbf{Eqn. 8} shows the sum over only relevant
\(\{\alpha\beta\mathbf{R}\}\) interactions to give the projected COHP
(pCOHP), which can be plotted onto the band structure or DOS.


\begin{equation}
\mathrm{iCOHP}_{\alpha\beta\mathbf{R}}
= \sum_{n,\mathbf{k}} \mathrm{COHP}_{\alpha\beta\mathbf{R}}^{n\mathbf{k}}\,f_{n\mathbf{k}}w_{\mathbf{k}}.
\tag{7}
\end{equation}

\begin{equation}
\mathrm{pCOHP}^{n\mathbf{k}}
= \sum_{\left\{\alpha\beta\mathbf{R}\right\}} \mathrm{COHP}_{\alpha\beta\mathbf{R}}^{n\mathbf{k}}.
\tag{8}
\end{equation}

With our additional \textbf{R} resolution, COGITO can filter
interactions not only by orbital type but by distance (even outside the
primitive cell). For example, the set of \(\{\alpha\beta\mathbf{R}\}\)
may include Si interactions only with its second nearest neighbors. This sets up the ability to analyze long-range bonding with more fidelity.

While the full nine-dimensional COHP is challenging to manage computationally,
our implementation vectorizes over all axes but \textbf{k}-point, 
balancing efficiency with memory requirements, and predetermines the indices of a 
user set \(\{\alpha\beta\mathbf{R}\}\) for pCOHP such that only the relevant slice of COHP is computed.

Finally, to analyze the bonding between atom A and atom B (in
\textbf{T}=\textbf{R} cell), we sum over orbitals belonging to the relevant atom.
This atom-level data informs the interactive crystal bond diagrams pictured in \textbf{Figs.~\ref{fig:vis_five}} and \textbf{\ref{fig:gan}}.  

\begin{equation}
\mathrm{iCOHP}_{AB\mathbf{ R}}
= \sum_{\alpha\in A,\beta\in B} \mathrm{iCOHP}_{\alpha\beta\mathbf{ R}}.
\tag{9}
\end{equation}

Similarly, a full nine-dimensional COOP can be created by replacing the Hamiltonian matrix with the overlap matrix.

\begin{equation}
\mathrm{COOP}_{\alpha\beta\mathbf{R}}^{n\mathbf{k}}
= {c_{\alpha n}^{\mathbf{k}}}^{\dagger}c_{\beta n}^{\mathbf{k}}\,S_{\alpha\beta}^{\mathbf{R}}\,
e^{i\mathbf{k}\cdot\left(\mathbf{R}-\boldsymbol{\tau}_{\alpha}+\boldsymbol{\tau}_{\beta}\right)}.
\tag{10}
\end{equation}

This COOP can be reduced in the same way as the COHP matrix in \textbf{Eqns. 7-9} but now indicates electron counts rather than electron energies. For example, the \(\mathrm{iCOOP}_{\alpha\beta\mathbf{R}}\) form provides an integrated electron charge analysis that reveals the quantity of electrons \textit{on} atomic orbitals and \textit{shared between} atomic orbitals. 
Then by summing over orbitals of each atom, as in Eqn. 9, the electrons can be partitioned \textit{on} atoms or \textit{between} atoms as bond charge, \(\mathrm{iCOOP}_{AB\mathbf{ R}}\). The electron count on atom A, \(\mathrm{q}_{A}^{onsite}\), is the special case where A=B and \textbf{R}=[0,0,0]. 

\begin{equation}
\mathrm{q}_{A}^{onsite}
= \mathrm{iCOOP}_{AA\mathbf{ 0}}.
\tag{11}
\end{equation}

However, since some electrons are in the bonds, summing \(\mathrm{q}_{A}^{onsite}\) over all atoms will not give the correct total electron count. 
If one wants all the electrons partitioned to atomic orbitals, rather than shared between atomic orbitals, summing all but one orbital axis of the iCOOP yields a charge-complete proxy for the electron occupation of each orbital.

\begin{equation}
\mathrm{q}_{\alpha}^{Mulli}
= \sum_{\beta,\mathbf{R}} \mathrm{iCOOP}_{\alpha\beta\mathbf{R}}.
\tag{12}
\end{equation}

Now, by summing \(\mathrm{q}_{\alpha}^{Mulli}\) over all orbitals of atom A and subtracting from the valence electron count, \(N_A\), of the atom's pseudo-potential, we arrive at the atomic partial charge from the Mulliken partitioning scheme.

\begin{equation}
\mathrm{q}_{A}^{Mulli}
= N_A - \sum_{\alpha\in A} \mathrm{q}_{\alpha}^{Mulli}.
\tag{13}
\end{equation}

This Mulliken-style partitioning can also be used for orbital-projected band and \textbf{k}-point resolved matrices, \(\mathrm{P}_{\alpha, n \textbf{k}}^{Mulli}\), which correctly normalizes to 1 when summing over \(\alpha\).

\begin{equation}
\mathrm{P}_{\alpha, n \textbf{k}}^{Mulli}
= \sum_{\beta,\mathbf{R}} \mathrm{COOP}_{\alpha\beta\mathbf{R}}^{n\mathbf{k}}.
\tag{14}
\end{equation}

\subsection{Computational details}

Structures for the benchmark set were taken from the dataset of
Vitale \textit{et al.}~\cite{vitale_automated_2020}. Structures for Si
(mp-149) and the GaN polymorphs (wurtzite, mp-804; sphalerite, mp-830;
hexagonal planar, mp-1007824; halite, mp-2853) were obtained from the
Materials Project~\cite{Jain2013}. VASP input files were
generated with pymatgen~\cite{Ong2013} from the \texttt{MPRelaxSet} defaults,
with the modifications described below.

All calculations used VASP~\cite{Kresse1996PRB,Kresse1996CMS} with the
PBE exchange--correlation functional~\cite{Perdew1996} and projector
augmented-wave (PAW) potentials~\cite{Blochl1994,Kresse1999} from the PBE.52
set. We used a plane-wave cutoff of 520~eV, \texttt{PREC = Accurate},
\texttt{LREAL = False} and \texttt{LASPH = True}. Calculations were
spin-unpolarized, and no Hubbard $U$ correction was applied. To reduce the
number of semicore states, the \texttt{MPRelaxSet} potentials were replaced
where possible: \texttt{\_pv} potentials by standard ones, and \texttt{\_sv}
potentials by \texttt{\_pv} potentials (or standard ones when no
\texttt{\_pv} version exists). The Brillouin zone was sampled with
$\Gamma$-centred meshes with a k-point spacing of 0.2~\AA$^{-1}$.

Each compound was calculated in three steps: (i) a full relaxation of atomic
positions and lattice vectors (\texttt{ISIF = 3}) using the
\texttt{MPRelaxSet} convergence criteria; (ii) a self-consistent static
calculation on the relaxed structure (\texttt{EDIFF} $= 10^{-5}$~eV,
Gaussian smearing of 0.01~eV), with
$N_\mathrm{bands} = 16N_\mathrm{atoms}$ to include enough
unoccupied states and saved wavefunctions (\texttt{LWAVE=True}); and (iii) a non-self-consistent
band structure from the static charge density, along the high-symmetry path
of Setyawan and Curtarolo~\cite{Setyawan2010}. COGITO (v0.3.13) with default settings was applied
to the output of step (ii), and its interpolated band structures were
compared with those from step (iii). 

\bibliography{COGITOrefs_new}

\end{document}